\documentclass[journal,9pt]{IEEEtran}
\usepackage{multirow}
\usepackage{booktabs}
\usepackage{amsmath,amsfonts}
\usepackage{algorithmic}
\usepackage{algorithm}
\usepackage{array}
\usepackage{comment}

\usepackage{float}
\usepackage[caption=false,font=normalsize,labelfont=sf,textfont=sf]{subfig}
\usepackage{amsmath}
\usepackage{commath}
\usepackage{enumitem}
\usepackage{bm}

\usepackage{textcomp}
\usepackage{stfloats}
\usepackage{url}
\usepackage{verbatim}
\usepackage{graphicx}
\usepackage{cite}
\usepackage[colorlinks=true, linkcolor=blue, citecolor=blue, filecolor=blue, urlcolor=blue]{hyperref}

\newcommand\blfootnote[1]{%
    \begingroup
    \renewcommand\thefootnote{}
    \footnotetext[0]{#1}
    \endgroup
}
\begin{document}

\title{\huge{Near-Field SAR Imaging of Moving Targets on Roads}}

\author{\Large{Raz Rajwan and Amir Boag, \textit{Fellow, IEEE}}}

\markboth{IEEE Transactions on Antennas and Propagation}%
{Shell \MakeLowercase{\textit{et al.}}: A Sample Article Using IEEEtran.cls for IEEE Journals}


\maketitle
\blfootnote{This work was supported in part by the BSF under Grant 2022346 and in 
 part by the NATO SPS under Grant G6118.
 
 The authors are with the School of Electrical and Computer Engineering, Tel Aviv University, Tel Aviv, 69978, Israel (e-mail: boag@tau.ac.il; rajwanraz@gmail.com).}

\begin{abstract}
This paper introduces a single-channel SAR algorithm designed to detect and produce high-fidelity images of moving targets in spotlight mode.    The proposed fast backprojection algorithm utilizes multi-level interpolations and aggregation of coarse images produced from partial datasets. Specifically designed for near-field scenarios and assuming a circular radar trajectory, the algorithm demonstrates enhanced efficiency in detecting both moving and stationary vehicles on roads.\end{abstract}


\section{Introduction}
\IEEEPARstart{S}{ynthetic}  Aperture Radar (SAR) has emerged as a powerful tool for remote sensing applications, offering high-resolution imaging capabilities regardless of weather conditions or time of day. SAR imaging is based on coherent processing of scattered signals over a range of frequencies and angles. The reflected signal can be modeled as a convolution between the pulse waveform and the ground reflectivity function \cite{Synthetic-Aperture-Radar-Signal-Processing-with-MATLAB-Algorithms}. A typical approach to SAR imaging involves the numerical evaluation of an integral operator \cite{Spherical_wave_near_field}. The integral is often formulated as a multidimensional Fourier transform allowing for efficient evaluation using Fast Fourier Transform (FFT) based algorithms\cite{BP_SAR_using_FFT} and \cite{SAR_image_formation_toolbox_for_MATLAB}. Utilizing the FFT lowers the computational cost of computing a two-dimensional $N\times N$ image from $O(N^4)$ to $O(N^2\log N)$, where $N$ denotes the number of samples in both the range and azimuth dimensions. However, the reduction of the imaging operator to FFT form is frequently achieved by far-reaching approximations, which cause degradation of the image quality. The common methods for SAR image formation include the Polar Format Algorithm (PFA) \cite{Spotlight_Synthetic_Aperture_Radar_Signal_Processing_Algorithms}, which operates under the assumption of the far-field approximation. Additionally, the Range Migration Algorithm (RMA) \cite{seismic_migration_techniques} presumes uniform sampling in both azimuth and range dimensions. The Chirp Scaling Algorithm (CSA) \cite{chirp_scaling} excludes interpolation but does not provide complete compensation for the spherical nature of the waves. A comprehensive comparison of these algorithms is given in \cite{Spotlight_Synthetic_Aperture_Radar_Signal_Processing_Algorithms} and \cite{Alg_and_implementation_computational_perspective}.

Several fast backprojection (BP) algorithms have been proposed in \cite{A_Boag_MLDD,N2logn_BP_alg,CBP_spotlight_SAR,Fast_Radon_transform,Fast_factorized_BP,A-Fast-Back-Projection-SAR-Imaging-Algorithm-Based-on-Wavenumber-Spectrum-Fusion-for-High-Maneuvering-Platforms}. These algorithms demonstrate greater imaging quality and flexibility when compared to the FFT-based ones, while retaining the same CC.

Here, we propose an improved algorithm for a single-channel radar by extending the divide-and-conquer approach proposed in \cite{A_Boag_MLDD} and \cite{BP_imaging_of_moving_objects_Agaibel}, with our focus on the near-field, spotlight mode scenario. In these papers, a recursive multi-level algorithm is introduced to efficiently compute the BP operator over the 4-D space. The adoption of the multi-level BP image formation is preferred due to its high
accuracy, robustness, and flexibility, particularly in compensating for variations in antenna
patterns across different angles and frequencies. Moreover, the BP algorithm has the capability to overcome near-field limitations that are often encountered in other algorithms.
In this study, we propose several modifications to improve the CC of the multi-level algorithm. Firstly, we identify roads within the static image. Subsequently, we demonstrate the
feasibility of implementing the multi-level algorithm solely on a 2-D space comprising the
detected roads, resulting in a significant reduction in complexity. Additionally, we adapt the multi-level algorithm to the near-field scenario and examine the limitations of
this method.

The article is organized as follows:
The BP operator for moving targets in the near field scenario is formulated in Sec. \ref{sec: BP formulation}. The 4-D multi-level algorithm and the adaptive algorithm for detecting and imaging moving targets are introduced in Sec. \ref{sec: fast algorithm formulation}. An improved algorithm for detecting and imaging targets on roads is presented in Sec. \ref{sec: Detect targets on roads}. The algorithm's performance is demonstrated and analyzed in Sec. \ref{sec: numerical analysis}. Finally, the conclusions are presented in Sec. \ref{sec: conclusion}.

\section{Problem Configuration and Imaging Operator}
\label{sec: BP formulation}
In this section, we describe the single channel SAR configuration. A circular radar trajectory is considered a practical choice, as shown in \cite{Moving-target-tracking-using-circular-SAR-imagery} and \cite{GMTI_reconstruction_in-single-channel-circular-SAR}. We assume that the radar moves along a circular trajectory of radius $R$ above a region of interest (ROI). The center of the radar trajectory is located above the center of the ROI.

\begin{figure}[H]
  \includegraphics[scale=0.7]{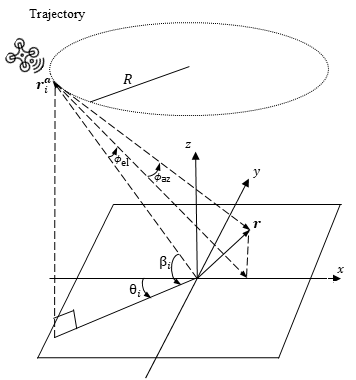}
  \centering
  \caption{System configuration.}
  \label{Sar Basic Configuration }
\end{figure}

\noindent In Fig \ref{Sar Basic Configuration }, $\beta_i$ represents the elevation angle, where $i$ is the slow time index. Here, $\mathbf{r}_{i}^a$ is the position of the antenna phase center at azimuth $\theta_{i}$, and $\mathbf{r}$ denotes a point on the $xy$-plane.
\begin{equation}\label{r vector}
\mathbf{r}=(x,y,0)
\end{equation}
The radar transmits at uniformly spaced angles $\theta_i$. Let $V_{t}(f_l)$ denote the Fourier transform of the transmitted signal, and $V_{r}(\theta_{i},f_{l})$ denote the Fourier transform of the received signal at an azimuthal angle $\theta_{i}$ $(i=1,...,N_{\text{p}}$), both at evenly spaced frequencies $f_l$ ($l = 1,...,N_f$). The normalized received signal data is represented as an $N_f \times N_{\text{p}}$ matrix

\begin{equation}\label{eq:example2}
P(\theta_{i},f_{l})=\frac{V_{r}(\theta_{i} ,f_{l}) } {V_{t}(f_{l})}
\end{equation}
and referred as the range profile matrix. We make the standard approximation that the radar does not move a significant distance while the pulse is being transmitted and received ("stop and go" assumption).
In a near-field scenario, the reflectivity distribution of the $xy$-plane, denoted  $g(\mathbf{r})$, can be evaluated using the discrete backprojection operator. Assuming planar and horizontal ROI, this operator is defined as:
\begin{equation}\label{eq: backprojection operator}
g(\mathbf{r}) = \sum_{i=1}^{N_{\text{p}}} \sum_{l=1}^{N_f}A(\mathbf{r},\mathbf{r}_i^\text{a},f_l)^{-1}P(\theta_{i},f_{l}) e^{j 2k_{l} \abs{\mathbf{r} - \mathbf{r}_i^\text{a}}}
\end{equation}

\noindent where $k_l=2\pi f_l/c$ is the wave number with $c$ being the speed of light. Also in \eqref{eq: backprojection operator}, a slowly varying factor $A$ is defined as
\begin{equation}\label{eq: Amplitude factor}
    A(\mathbf{r},\mathbf{r}_i^\text{a},f_l) = \frac{F^2(\bm{\phi},f_l)}{\abs{\mathbf{r}-\mathbf{r}_i^\text{a}}^2}
\end{equation}
where $\abs{\mathbf{r}-\mathbf{r}_i^\text{a}}^2$ accounts for the two-way propagation loss. Also in \eqref{eq: Amplitude factor}, $F$ represents the field pattern of the radar antenna, which depends on the frequency $f_l$
and the observation direction $\bm{\phi} = (\phi_{\text{el}},\phi_{\text{az}})$. This observation direction is determined by the radar position, $\mathbf{r}_i^\text{a}$, and the imaging point $\mathbf{r}$. From this point on, we assume $N_f=N_{\text{p}}=N$. The reflectivity distribution is evaluated on a discrete, evenly spaced $N \times N$ grid of spatial locations $\mathbf{r}$. Brute-force computation of the reflectivity using the discrete form of the BP operator (\ref{eq: backprojection operator}) is characterized by $O(N^4)$ complexity.\\

Equation \eqref{eq: backprojection operator} can be modified to allow imaging of targets moving with constant velocity $\mathbf{v}=(v_x,v_y,0)$, as shown in \cite{BP_imaging_of_moving_objects_Agaibel}:
 
 \begin{equation}\label{eq: backprojection operator dynamic}
g(\mathbf{r},\mathbf{v})=\sum_{i=1}^{N}\sum_{l=1}^{N}A(\mathbf{r},\mathbf{r}_i^\text{a},f_l)^{-1}P(\theta_{i},f_{l})e^{j2 k_{l}\abs{(\mathbf{r}+\mathbf{v}t_i)-\mathbf{r}_i^\text{a}}}
\end{equation}

\noindent where $t_i$ is the slow time corresponding to $\theta_i$. Directly evaluating \eqref{eq: backprojection operator dynamic} on an $N^4$-point spatial location and velocity grid ($\mathbf{r},\mathbf{v}$) would require staggering $O(N^6)$ operations. 
To that end, the following section introduces algorithms designed to reduce the CC.

\section{Fast Algorithms}
\label{sec: fast algorithm formulation}
\subsection{4-D Multi-Level Imaging Algorithm}\label{subsec: MLDD}
A recursive multi-level algorithm, originally introduced in \cite{A_Boag_MLDD} for the static case, was adapted to the dynamic case through appropriate modifications described in \cite{BP_imaging_of_moving_objects_Agaibel}. This algorithm, known as the multi-level domain decomposition (MLDD) method, is further enhanced in this work by proposing an iterative implementation.

In the initial phase, the input data is partitioned into subdomains of size \( N_{\text{c}} \times N_{\text{c}} \). For each subdomain, a low-resolution image is computed using expression \eqref{eq: backprojection operator dynamic} on a discrete grid defined by \((\mathbf{r}^{L_1}, \mathbf{v}^{L_1})\). At the first resolution level, $L_1$, defined as $\log_2(N_{\text{c}})$, the number of grid points in each dimension is $N_{\text{c}}$, while at the final resolution level, \( L_{\text{max}} \), defined as $\log_2(N)$, the number of grid points is $N$. Thus at level $L_1$, instead of computing pixel values at \( N \) points per dimension, the evaluation is performed at only $N_{\text{c}}$ points, significantly reducing the computational cost. This results in \((N/N_{\text{c}})^2\) level-$L_1$ low-resolution images, which serve as a starting point for further refinement at higher resolution levels.

In the second stage, groups of four low-resolution images based on adjacent data subdomains are repeatedly interpolated and aggregated to generate finer resolution images. This process is performed iteratively, each iteration increasing the resolution level by one, progressing from $L_1$ to $L_\text{max}$. Prior to interpolation, it is required to eliminate rapid oscillations introduced by the exponential term in \eqref{eq: backprojection operator dynamic}. A phase correction is applied to shift the center frequency to zero, making the term amenable to interpolation. This interpolation process doubles the number of grid points for each dimension. After interpolation, the original phase is restored. The iterative process of interpolation and aggregation continues until a high-resolution image, $\Tilde{g}^{L_{\text{max}}}_{11}(\mathbf{r}^{L_\text{max}},\mathbf{v}^{L_\text{max}})$ is obtained at the final level, $L_\text{max}$. For simplicity, from this point on, the $pq$th low-resolution image at level $L$ will be referred to as $\Tilde{g}^{L}_{pq}$ omitting the explicit grid notation.

In Algorithm \ref{alg: 2D adaptive multi level algorithm}, a pseudo code illustrates the two stages described above. Here, $P_{pq}, k_p, \mathbf{r}^a_q$ and $t_q$ are the range profile, wave number, antenna location, and slow time in the $pq$th data subdomain, respectively. The symbol $\circ$ denotes the Hadamard (element-wise) multiplication. The asterisk represents the element-wise conjugate operator. 
The operator $I_{x y v_x v_y}^{L-1}$ performs a 4-dimensional interpolation to refine the resolution of the grid. It transforms the grid from $(\mathbf{r}^{L-1}, \mathbf{v}^{L-1})$ to a higher-resolution grid $(\mathbf{r}^{L}, \mathbf{v}^{L})$, where each dimension contains $2^L$ spatial locations and velocities. Linear interpolation exhibits good target detection performance with a low CC. Prior to the interpolation, the $pq$th low-resolution image is multiplied by the conjugate of the exponential term $E_{pq}^{L-1} = \exp(j2\Bar{k_{p}}\abs{\mathbf{r}+\mathbf{v}\Bar{t_{q}}-\Bar{\mathbf{r}}_{q}^a})$ to mitigate phase variation and facilitate interpolation. Here, $\Bar{k_{p}}$, $\Bar{\mathbf{r}}_{q}^a$, and $\Bar{t_{q}}$ represent the average wave number, antenna location, and slow time for the $pq$th data subdomain, respectively. Following interpolation, the exponent term, $E_{pq}^{L}$, is multiplied to restore the correct phase. In this context, the parameter $L_\text{d}$ is set to $L_{\text{max}}$ and will be explained in the next section.

The CC at the final level is $O(N^4)$ due to the necessity of performing $O(N)$ operations for each output dimension. At the previous level, the number of operations is reduced by half for each output dimension, resulting in a decrease by a factor of 16, while simultaneously doubling the number of low-resolution images for each input variable, leading to a total decrease by a factor of 4. Consequently, the complexity is reduced by a factor of 4 for the preceding stages, thus behaving like a geometric regression. As a result, the CC of the algorithm is $O(N^4)$.

In addition to its high complexity of $O(N^4)$, this algorithm requires substantial memory resources. Storing an $N^4$-element array at the final level becomes impractical as the value of $N$ increases. The following algorithm presents a methodology for reducing both computational complexity and memory demands.
\subsection{Adaptive Detection and Imaging} \label{subsec: AMLDD}
An adaptive version of this algorithm, referred to as the 
 4-D adaptive multi-level domain decomposition (AMLDD) method, was proposed in \cite{BP_imaging_of_moving_objects_Agaibel}. The algorithm follows the same initial steps as the original MLDD but introduces a predefined detection level, denoted as $L_{\text{d}}$. This detection level is selected as a trade-off between CC and detection performance. Specifically, the detection level, $L_\text{d}$, is constraint to the range $\log_2(N_\text{c})\leq L_\text{d}\leq\log_2(N)$. Once the detection level is reached, an incoherent summation of the intermediate images is performed as:
\begin{equation}\label{eq: Detection matrix}
    D^{L_{\text{d}}}(\mathbf{r}^{L_{\text{d}}},\mathbf{v}^{L_{\text{d}}}) = \sum_{p=1}^{N/2^{L_{\text{d}}}}\sum_{q=1}^{N/2^{L_{\text{d}}}}\abs{g_{pq}^{L_{\text{d}}}}
\end{equation}
For simplicity, this detection matrix will be referred to as $D$
from this point on, omitting subscripts and grid notation. The detection matrix, $D$, is a four-dimensional matrix, with each dimension consisting of $2^{L_\text{d}}$ elements. Local maxima in $D$ represent static or moving targets on a coarse grid $(\mathbf{r}^{L_\text{d}},\mathbf{v}^{L_\text{d}})$. Searching for these local maxima in $D$ requires the following computational cost:
\begin{equation}\label{eq:computational cost}
O\left(\left(\frac{N}{2^{L_{\text{d}}}}\right)^2 \left(2^{L_{\text{d}}}\right)^4\right) = O\left(N^2 2^{2L_{\text{d}}}\right)
\end{equation}
Setting the detection level at the middle level, $L_{\text{d}}=L_{\text{max}}/2=\log_2(N)/2$, results in a search complexity of $O(N^3)$. These stages are illustrated in Algorithm \ref{alg: 2D adaptive multi level algorithm}.
After detecting the targets on a broad, coarse grid, we can enhance the precision of both the location and speed. This is done by continuing to interpolate and aggregate the images to the final level within a local, high-resolution grid centered around the initially detected target. The full resolution calculation, assuming a bounded number of small detected regions, requires $O(N^2)$ operations. After completing the high-resolution calculations, we proceed to clean the image from the smearing induced by moving targets.

\subsection{Image Cleaning} \label{subsec: image cleaning}
Moving targets appear defocused and smeared in the static image. Our aim is to "clean" the static image by eliminating the moving targets influence from the input matrix $P$. A full cleaning procedure is proposed in \cite{BP_imaging_of_moving_objects_Agaibel} for far field scenario. This cleaning method is adapted for the near field scenario with the CC of $O(N^2)$ for each moving target.

\begin{figure}[ht]
\centering
\subfloat[]{\includegraphics[width=0.24\textwidth]{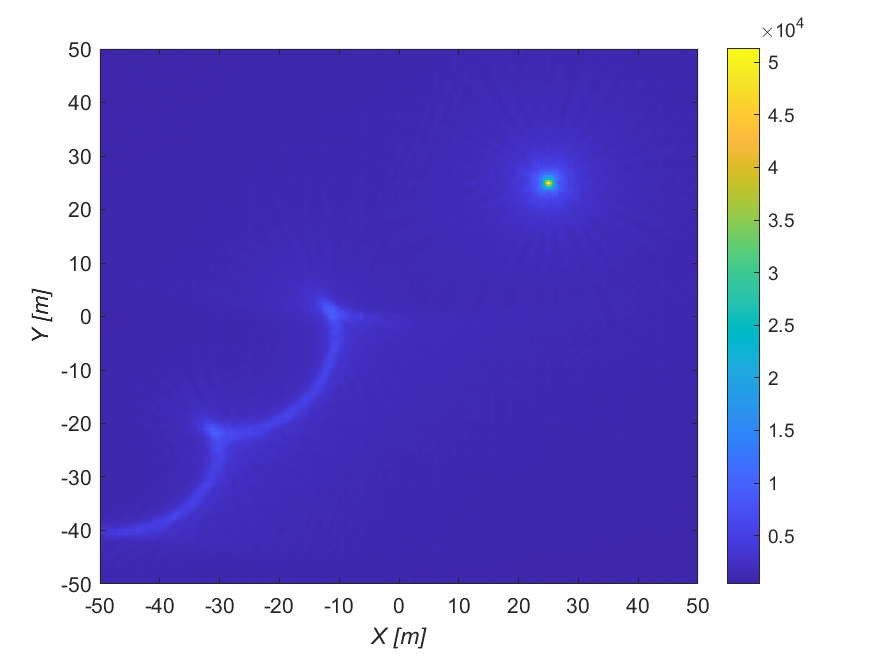}%
\label{fig: full image 4-d (a)}}
\hfil
\subfloat[]{\includegraphics[width=0.24\textwidth]{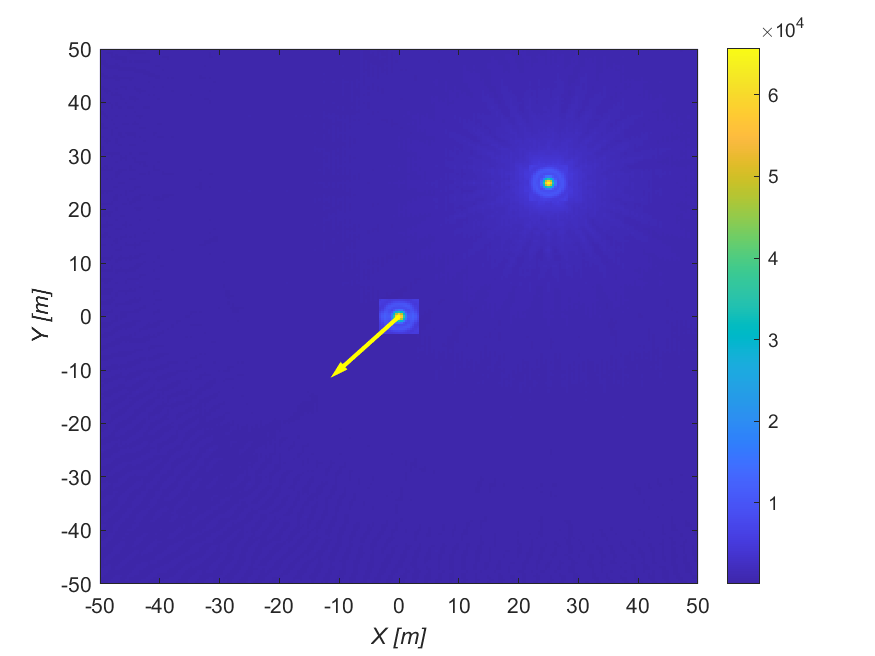}%
\label{fig: full image 4-d (b)}}
\caption{Scenario 1 comprising a static target and a moving target (a) Static image (b) Static image after detecting and cleaning moving targets.}
\label{fig: full image 4-d}
\end{figure}

Fig. \ref{fig: full image 4-d} illustrates the full imaging procedure. The detection matrix, $D$, is computed using the adaptive algorithm and then searched for local maxima. A resolution upgrade is applied in the detected target's area, and moving targets are removed from the range profile matrix, $P$. Following this, the stationary image is constructed using one of the algorithms with the CC of $O(N^2\log N)$. Finally, incorporating a directional indicator highlighting the initial position and velocity for each of the moving targets.

\noindent The next section introduces further improvements to the algorithm, with the aim of reducing the CC below O($N^3$) by incorporating road characteristics. 
\section{Target detection on roads}
\label{sec: Detect targets on roads}
\subsection{Road Detection} \label{subsec: road detection}
The primary goal of road detection is to identify key road features, such as the starting point and orientation of each road, within a static image. These features allow for a significant reduction in computational workload by limiting the application of the BP operator to the identified road segments, rather than the entire space and velocity grid ($\mathbf{r}, \mathbf{v}$). This in turn enhances the efficiency of the adaptive multi-level algorithm and represents the main contribution of this work.
\begin{figure}[ht]

\subfloat[]{\includegraphics[width=0.24\textwidth]{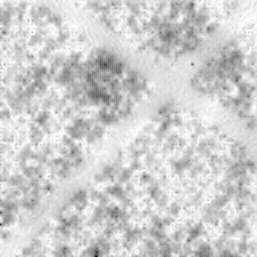}\label{fig: Image Processing steps(a)}}\hfil
\subfloat[]{\includegraphics[width=0.24\textwidth]{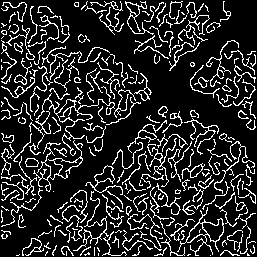}\label{fig: Image Processing steps(b)}}\vfill%
\subfloat[]{\includegraphics[width=0.24\textwidth]{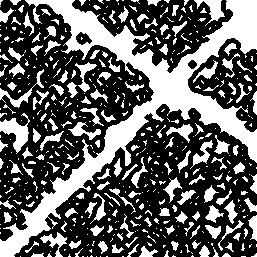}\label{fig: Image Processing steps(c)}}\hfill%
\subfloat[]{\includegraphics[width=0.24\textwidth]{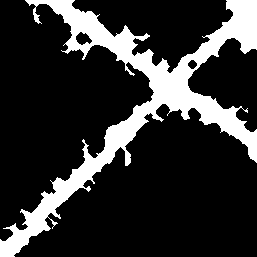}\label{fig: Image Processing steps(d)}}%
\caption{Image processing steps; (a) Gray scale image (b) Edge filter (c) Dilation (d) Connected components.}
\label{fig: Image Processing steps}
\end{figure}
\noindent The following steps are used to detect roads in the static image, which is generated by applying one of the stationary imaging algorithms to the input data.

\begin{enumerate}[label=\alph*)]
    
    \item The pixel values of the static image are normalized to the range [0, 1].
    \item An edge image is created using the Canny filter \cite{An-improved-Canny-edge-detection-algorithm}, \cite{Using-the-Canny-edge-detector-for-feature-extraction...}.
    \item The image undergoes dilation with a disk-shaped structural element, expanding the detected edges to bridge gaps and improve the continuity of road segments.
    \item Connected components \cite{Connected_Components_Labeling} are identified in the dilated image to highlight road segments. Small-area connected components are then filtered out, preserving only the road segments.
\end{enumerate}

Fig. \ref{fig: Image Processing steps} illustrates the image processing steps employed for segmenting the roads in the static image. The extraction of the road parameters is shown in the following steps:
\begin{enumerate}[label=\alph*)]
    \item Perform the Hough transform on the connected components image. The Hough transform maps points in the image to a parameter space, where straight lines are represented as peaks corresponding to potential line segments. Using the Hough transform for road detection has been shown as an effective method in \cite{Road_extraction_from-high-resolution-remote_sensing_images_using_wavelet_transform_and_hough_transform,Road_Detection_by_Using_a_Generalized_Hough_Transform,Road_detection_from_aerial_imagery}.
    \item Detect peaks in the transformed image. The number of peaks exceeds the number of roads. Each road may consist of multiple lines.
    \item Extract line parameters ($\rho_n, \alpha_n$) from the Hough transform, where $\rho_n$ represents the distance from the origin to the line of interest, and $\alpha_n$ defines the angle of the line relative to the $x$-axis. 
    \item To determine the number of roads and their parameters, including starting points and directions, we employ an unsupervised clustering algorithm known as mean shift clustering \cite{Clustering_Analysis_Based_on_the_Mean_Shift}. 
    \item Convert the pixel locations of the image to a global Cartesian coordinate system. 
\end{enumerate}

\begin{figure}[ht]
\subfloat[]{\includegraphics[width=0.14\textwidth]{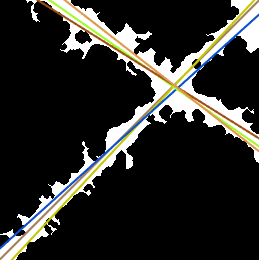}\label{fig: Parameters Extraction(a)}}\hfil%
\subfloat[]{\includegraphics[width=0.20\textwidth]{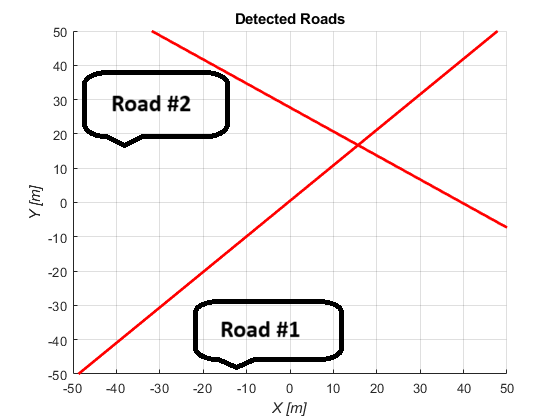}\label{fig: Parameters Extraction(b)}}\vfill%
\caption{Parameter extraction: (a) Image with detected lines; (b) Road parameters: $(\rho_1,\alpha_1)$ = $(0,45^\circ)$ and $(\rho_2,\alpha_2)$ = $(25,-35^\circ)$.}
\label{fig: Parameters Extraction}

\end{figure}
The resulting Hough transform lines are depicted in Fig. \ref{fig: Parameters Extraction(a)}. Lines that are adjacent and belong to the same road are centralized and grouped together using mean shift clustering. The lines extracted from the static image are illustrated in Fig. \ref{fig: Parameters Extraction(b)}. 
Enhancing the road segmentation and parameter extraction could be achieved through deep learning techniques; however, this falls beyond the scope of the current paper. To leverage the extracted road parameters for more efficient target detection, we introduce a road-based Algorithm that minimizes the computational complexity by focusing on already identified road segments.
\subsection{Road-based Algorithm} \label{subsec: Road Based algorithm}
With the road parameters identified, we can reduce the dimensionality of the output grid by aligning the image such that each road runs along the $x$-axis. This is achieved by applying the following transformation to the antenna phase center coordinates:
\begin{equation}\label{Rotation equation}
\mathbf{r}^\text{a}_{i,\text{rot}} = 
\begin{bmatrix}
\cos\alpha & -\sin\alpha\\
\sin\alpha & \cos\alpha \\
\end{bmatrix} \left[
\begin{bmatrix}
x_i^{\text{a}} \\
y_i^{\text{a}} \\
\end{bmatrix}- \begin{bmatrix}
0 \\
\frac{\rho}{\sin\alpha} \\
\end{bmatrix}\right]
\end{equation}
In this equation, the first matrix is the standard 2-D rotation by $\alpha$, $[x_i^{\text{a}}\; y_i^{\text{a}}]^T$ represents the antenna phase center coordinates $\textbf{r}_i^\text{a}$, and the term $\rho/\sin\alpha$ ensures that the road passes through the origin before the rotation is applied.
 This process is repeated for each identified road within the static image. For each rotated image, we execute the 2-D road-based algorithm described in Algorithm \ref{alg: 2D adaptive multi level algorithm}, setting the output arguments to $\mathbf{r} = x$ and $\mathbf{v} = v_x$. The output grid, $(x, v_x)$, has dimensions $N \times N$, and the output detection matrix, $D$, is of size $2^{L_\text{d}} \times 2^{L_\text{d}}$.

\begin{algorithm}[H]
\caption{Imaging Algorithm} \label{alg: 2D adaptive multi level algorithm}
\begin{algorithmic}
\STATE 

\STATE \textbf{PROCEDURE} Imaging Alg ($P,k,t,\mathbf{r}^\text{a},\mathbf{r},\mathbf{v},N_{\text{c}},N,L_{\text{d}}$)
\STATE [First stage: calculation of low-resolution images] 
\STATE \textbf{for} $p\text{, }q=1$ to $N/N_{\text{c}}$ 

\STATE \hspace{0.5cm}$ \Tilde{g}_{pq}^{L_1} = \sum_{i=1}^{N_{\text{c}}}\sum_{l=1}^{N_{\text{c}}}P_{pq}(i,l) e^{j 2 k_p(l) \abs{(\mathbf{r}^{L_1}+\mathbf{v}^{L_1}t_q(i))-\mathbf{r}_q^a(i)}}$
\STATE \textbf{end for}
\STATE [Second stage: interpolation and aggregation of adjacent images]

\STATE \textbf{for} $L=L_1+1$ to $L_\text{d}$
\STATE \hspace{0.5cm}\textbf{for} $p\text{, }q=1$ to $N/2^L$
\STATE \hspace{1cm}$\Tilde{g}_{pq}^{L}= 0$
\STATE \hspace{1cm}\textbf{for} $a=1\text{, }b=1$ to 2
\STATE \hspace{1.5cm}$p'= 2(p-1)+a$
\STATE \hspace{1.5cm}$q'= 2(q-1)+b$
\STATE \hspace{1.5cm}$\Tilde{g}_{pq}^{L} = \Tilde{g}_{pq}^{L} + E_{p'q'}^L\circ I_{x y v_xv_y}^{L-1}(E_{p'q'}^{L-1})^*\circ\Tilde{g}_{p'q'}^{L-1}$
\STATE \hspace{1 cm}\textbf{end for}
\STATE \hspace{0.5 cm}\textbf{end for}
\STATE \textbf{end for}
\STATE \textbf{$D=0$ }
\STATE \textbf{for} $p\text{, }q=1$ to ${N/2^{L_{\text{d}}}}$
\STATE \hspace{0.5cm} $D=D+\abs{\Tilde{g}_{pq}^{L_{\text{d}}}}$
\STATE \textbf{end for}
\STATE \textbf{return} $D,\Tilde{g}$
\STATE \textbf{end PROCEDURE}

\end{algorithmic}
\end{algorithm}
Given that the road is aligned with the $x$-axis, the search is optimized as there is no need to search along the $y$ and $v_y$ axes, resulting in a significant reduction in the CC. Consequently, the maxima search process within the detection matrix, $D$, is also improved, as $D$ now has two dimensions instead of four. Upon detecting all targets along the specific road, we improve their resolution and revert the antenna phase center coordinates to their original position. This process is repeated for each road, leading to $O(N^2)$ operations per road. Howev
er, if the road has a non-negligible width, then the algorithm must be applied multiple times at different $y$-values to cover the entire cross-section of the road and ensure that all potential targets are detected. This repetition does not change the overall complexity of the method.

\begin{figure}[H]
\begin{center}
    \subfloat[]{\includegraphics[width=0.4\textwidth]{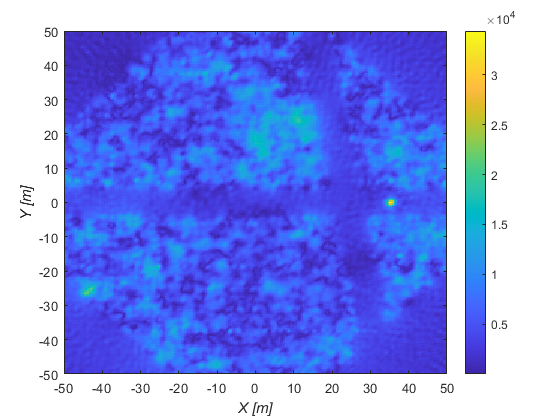}\label{fig: Rotated Static image(a)}}
\end{center}
\caption{Rotated static image. One road is aligned with the $x$-axis.}
\label{fig: Rotated Static image}
\end{figure}
The pseudo code in Algorithm \ref{alg:Pseudo algorithm} illustrates the full imaging procedure. Initially, a static image is generated utilizing the $O(N^2\log N)$ algorithm (MLDD). Subsequently, the road detection algorithm identifies roads within the image and extracts their parameters. For each road, the coordinates of the antenna phase center are shifted and rotated, and the 2-D adaptive algorithm is executed solely on the $x$ and $v_x$ dimensions. Then, local maxima in $D$ are identified. Target's resolution is upgraded by developing the image to $L_{\text{max}}$ at the detected areas. Subsequently, the antenna coordinates and the coordinates of located targets are rotated back. Finally, any moving targets are cleaned from the range profile matrix, $P$, and the final image is synthesized.
\begin{algorithm}[H]
\caption{Full Imaging Procedure} \label{alg:Pseudo algorithm}

\begin{algorithmic}
\STATE
\STATE \textbf{PROCEDURE} Road-based Alg ($P, k ,t ,\mathbf{r}^\text{a} ,\mathbf{r} ,\mathbf{v} ,N_{\text{c}} ,N,L_{\text{d}}$)

\STATE $\tilde{g}$ = \hyperref[alg: 2D adaptive multi level algorithm]{Imaging Alg}($P,k,t,\mathbf{r}^\text{a},\mathbf{r},\textbf{v} = 0,N_{\text{c}},N,L_{\text{max}}$) [static image]
\STATE$(\rho_n,\alpha_n)$ = \hyperref[subsec: road detection]{Road Detection}($\tilde{g}$)
\STATE \textbf{for} each road:
\STATE  \hspace{0.5 cm} $\mathbf{r}^\text{a}_\text{rot}$ 
  = \hyperref[Rotation equation]{Rotate}($\mathbf{r}^\text{a},\rho_n$,$-\alpha_n$) 
\STATE  \hspace{0.5 cm} $D$ 
 = \hyperref[alg: 2D adaptive multi level algorithm]{Imaging Alg}($P,k,t,\mathbf{r}^\text{a}_\text{rot},x,v_x, N_{\text{c}},N,L_{\text{d}}$)
\STATE  \hspace{0.5 cm} $T$ = 
 Find Local Maxima($D$)
\STATE  \hspace{0.5 cm} $T$ = Upgrade Resolution($T$)

\STATE  \hspace{0.5 cm} Detected Targets += T
\STATE \textbf{end for}
\STATE $P$ = \hyperref[subsec: image cleaning]{Image Cleaning}($P$, Detected Targets)
\STATE $\tilde{g}$ = 
 \hyperref[alg: 2D adaptive multi level algorithm]{Imaging Alg}($P,k,t,\mathbf{r}^\text{a},\mathbf{r},\textbf{v} = 0,N_{\text{c}},N$)
\STATE $\tilde{g}$ = Add Target Indicators($\tilde{g}$, Detected Targets)
\STATE \textbf{return} $\tilde{g}$
\STATE \textbf{end PROCEDURE}
\end{algorithmic}
\end{algorithm}

\section{Numerical results}
\label{sec: numerical analysis}
\begin{figure}[H]
\begin{center}
    {\includegraphics[width=0.4\textwidth]{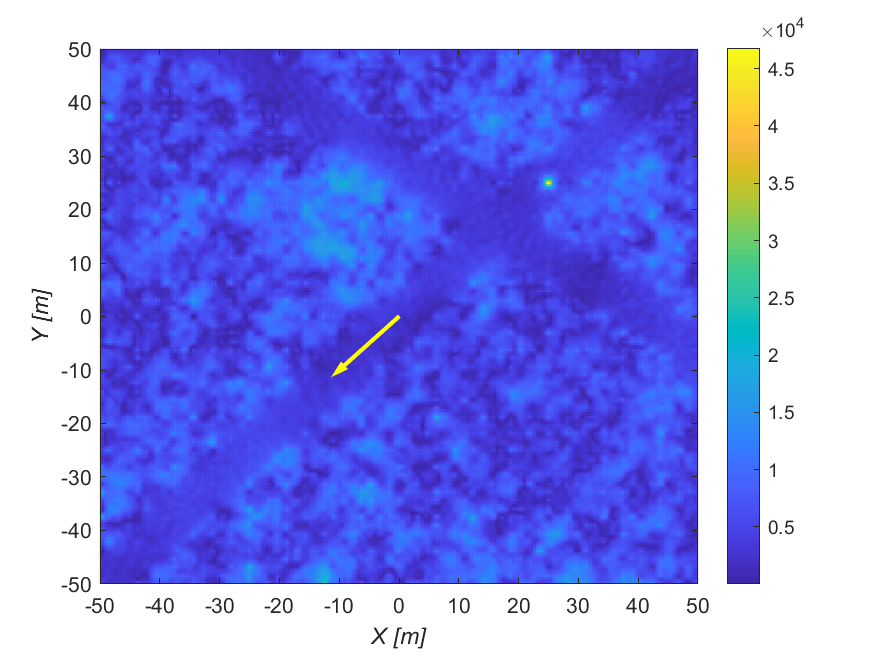}}
\end{center}
\caption{Final image with clutter. }
\label{fig: final image clutter}
\end{figure}

\begin{figure}[H]
\begin{center}
    {\includegraphics[width=0.48\textwidth]{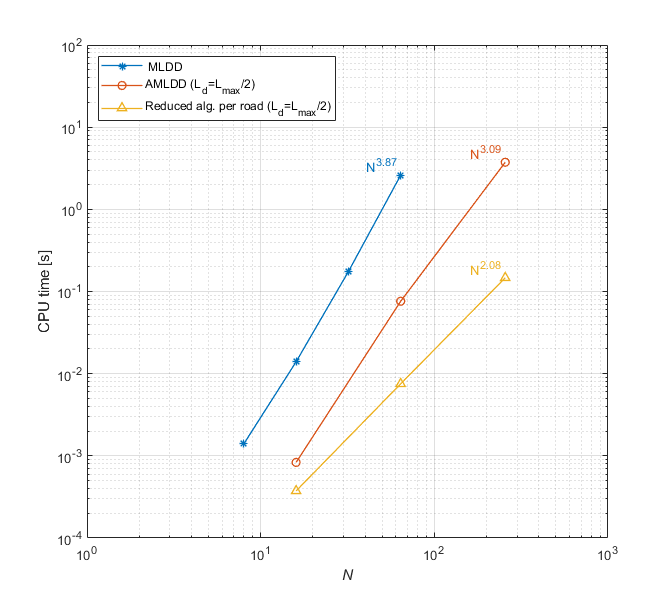}}
\end{center}
\caption{Run time of the described algorithms.  }
\label{fig: Run time comparison}
\end{figure}

The findings presented in this section pertain to a carrier frequency of 3 GHz, with a range resolution of 4 m. The radar operates at an altitude of 200 m and follows a circular trajectory with a radius of 200 m, $N$ = 256. The scenario involves the detection of two distinct targets, Target 'a' is stationary at coordinates $(25,25)$, whereas target 'b' starting its movement from $(0,0)$ with a constant velocity of $(-0.16,-0.16)$ m$/\Delta t$, where $\Delta t$ represents the time interval between successive pulses.
In this situation, there are two types of clutter: one inside the roads and one outside. The clutter outside the roads is stronger.
Following the procedure outlined in Algorithm \ref{alg:Pseudo algorithm}, we first extract the road parameters utilizing the static image. For every road detected, we adjust the image orientation such that the road aligns parallel to the $x$-axis (see Fig. \ref{fig: Rotated Static image}). Subsequently, we execute the road-based algorithm, with initial block size of $N_{\text{c}}=8$. In the initial stage, we evaluate (\ref{eq: backprojection operator dynamic}) at $N_{\text{c}}$ points, for each of the output dimensions ($x$ and $v_x$). Then, we interpolate and aggregate the intermediate images until we reach the detection level $L_{\text{d}}=L_{\text{max}}/2=4$. The final image (after detecting and cleaning moving targets) is depicted in Fig. \ref{fig: final image clutter}. 
A theoretical comparison of the CCs of the discussed algorithms is presented in Table \ref{tab:CC table}. Fig. \ref{fig: Run time comparison} presents comparison among of the full 4-D, the adaptive, and the road-based algorithms. The theoretical CCs of the simulated algorithms are confirmed.

\begin{table}[H]
 \caption{CC of the described algorithms}
  \label{tab:CC table}
  \centering
  \begin{tabular}{|c|c|}
    \hline
    \textbf{Algorithm} & \textbf{Computational time}\\
    \hline
    
    Direct backprojection & $O(N^6)$  \\
    \hline
    FFT-based backprojection & $O(N^4\log N)$ \\
     \hline
    \hyperref[subsec: MLDD]{4-D algorithm} ($L_{\text{d}}=L_{\text{max}}$)& $O(N^4)$  \\
     \hline
    \hyperref[subsec: AMLDD]{Adaptive 4-D algorithm} ($L_{\text{d}}=L_{\text{max}}/2$) & $O(N^3)$\\
    \hline
    \hyperref[subsec: Road Based algorithm]{Road-based algorithm} ($L_{\text{d}}=L_{\text{max}}/2$) & $O(N^2)$ per road\\
    \hline
  \end{tabular}
\end{table}

To evaluate the detection performance of the method in the presence of clutter, we simulate two scenarios with clutter (dynamic and static targets) by placing one point scatterer with a random phase in each resolution cell of area $\Delta x \Delta y$, where $\Delta x$ and $\Delta y$ are the spatial sampling intervals. The SCR is defined as the ratio $\sigma_t/\sigma_c$, where $\sigma_t$ is the target radar cross section (RCS) and $\sigma_c = \sigma_0 A_c$ is the RMS clutter RCS, $\sigma_0$ being the clutter scattering coefficient and $A_c = \Delta x \Delta y$ is the clutter area. The SCR level of 0 dB is determined by equalizing the target's image peak value to the clutter RMS response, both on the final stationary image. Table \ref{tab: SCR for detection} illustrates the SCR observed in the detection arrays as a function of the static SCR for multiple detection levels, $L_{\text{d}}$.

\begin{table}[ht!]
\centering
\caption{SCR for Detection Arrays at Different Levels}
\label{tab: SCR for detection}
\begin{tabular}{|c|c|c|c|c|c|}
\hline
\textbf{Static} & \textbf{Target} & \multicolumn{4}{c|}{\textbf{SCR at target’s velocity (dB)}} \\
\cline{3-6}
\textbf{SCR} & & \multicolumn{4}{c|}{\textbf{Adaptive method}} \\
\cline{3-6}
& & $L_{\text{d}}=4$ & $L_{\text{d}}=5$ & $L_{\text{d}}=6$ & $L_{\text{d}}=7$ \\
\hline
\multirow{2}{*}{19 dB} & static & 1.87 & 4.06 & 6.86 & 9.99 \\
\cline{2-6}
& dynamic & 2.01 & 4.12 & 7.51 & 10.7 \\
\hline
\multirow{2}{*}{26 dB} & static & 4.36 & 7.32 & 10.4 & 13.2 \\
\cline{2-6}
& dynamic & 4.52 & 7.85 & 10.7 & 13.5 \\
\hline
\end{tabular}
\end{table}

\section{Conclusion}
\label{sec: conclusion}
In summary, our paper introduces a novel algorithm employing the BP approach specifically designed to detect and image targets moving along roads in near-field scenarios. This algorithm efficiently computes high-resolution SAR images, requiring $O(N^2)$ operations per road in the image. Additional $O(N^2\log N)$ operations are required to form the static image. Further speedup of the proposed algorithm was easily achieved by processing the coarse images
 in parallel through the use of multi-threading. Optimal detection
 performance is obtained through linear interpolation, which necessitates a minimum of
 8 dB of static SCR and a high detection level. As SCR improves, it becomes feasible
 to employ lower detection levels, thereby reducing computational time.

\bibliographystyle{IEEEtran}

\vfill

\end{document}